\documentclass[runningheads]{svmult}

\usepackage{makeidx}   
\usepackage{graphicx}  
\usepackage{subeqnar}  
\usepackage{multicol}  
\usepackage{physprbb}  
\makeindex             

\newcommand{\be}{\begin{equation}}\newcommand{\ee}{\end{equation}}
\newcommand{\bea}{\begin{eqnarray}}\newcommand{\eea}{\end{eqnarray}}
\newcommand{\brr}{\begin{array}}\newcommand{\err}{\end{array}}
\newcommand{\bit}{\begin{itemize}}\newcommand{\eit}{\end{itemize}}
\newcommand{\ben}{\begin{enumerate}}\newcommand{\een}{\end{enumerate}}
\newcommand{\ba}{\begin{array}}
\newcommand{\ea}{\end{array}}
\def\lab{\label}\def\lan{\langle}
\def\lf{\left}

\def\non{\nonumber}\def\pa{\partial}\def\ran{\rangle}

\def\ri{\right}
\def\ga{\gamma}\def\Ga{\Gamma}

\def\ka{\kappa}

\def\Om{\Omega}

\def\1{{_{1}}}\def\2{{_{2}}}

\newcommand{\IJ}{{_I}}
\newcommand{\II}{{_{I\!I}}}

\begin{document}
\title*{Dissipation, Emergent Quantization and Quantum Fluctuations}
\toctitle{Dissipation, Emergent Quantization and Quantum
Fluctuations}
%
%
\titlerunning{Dissipation, Emergent Quantization and Quantum Fluctuations}
%
\author{Massimo Blasone\inst{1}\inst{2}
\and Petr Jizba\inst{3} \and Giuseppe Vitiello\inst{2}}
\authorrunning{M.Blasone, P.Jizba and G.Vitiello}
%
%
\institute{Theoretical Physics, Blackett Laboratory,
Imperial College London, SW7 2AZ London, U.K. \and Dipartimento di  Fisica
"E.R.Caianiello", INFN and INFM, Universit\`a di Salerno, I-84100
Salerno, Italy \and Institute of Theoretical Physics, University
of Tsukuba, Ibaraki 305-8571, Japan }

\maketitle              

\begin{abstract}
We review some aspects of the quantization of the damped harmonic
oscillator. We derive the exact action for a damped mechanical
system in the frame of the path integral formulation of the
quantum Brownian motion problem developed by Schwinger and by
Feynman and Vernon. The doubling of the phase-space degrees of
freedom for dissipative systems and thermal field theories is
discussed and the doubled variables are related to quantum noise
effects. The 't Hooft proposal, according to which the loss of
information due to dissipation in a classical deterministic system
manifests itself in the quantum features of the system, is
analyzed and the quantum spectrum of the harmonic oscillator is
shown to be originated from the dissipative character of the
original classical deterministic system.
\end{abstract}

\section{Introduction}
Our purpose in this report is to derive the action for a damped
mechanical system in the path integral formalism, to discuss the
role of quantum fluctuations and to show that the loss of information
due to dissipation manifests itself in the form of quantum noise
effects.

A microscopic theory for a dissipative system must include the
details of the processes responsible for dissipation. One
would then start with a Hamiltonian that describes the system, the bath
and the system-bath interaction. The description of the original
dissipative system is recovered by the reduced density matrix
obtained by eliminating the bath variables which originate the
damping and the fluctuations. In quantum mechanics canonical
commutation relations are not preserved by time evolution due to
damping terms. The role of fluctuating forces is in fact the one
of preserving the canonical structure. However, the knowledge of
the details of the processes inducing the dissipation may not
always be possible; these details may not be explicitly known and
the dissipation mechanisms are sometime globally described by such
parameters as friction, resistance, viscosity etc.. In some sense,
such parameters are introduced in order to compensate the
information loss caused by dissipation.

Our discussion in the present paper is aimed to consider, from one
side, the description of dissipative systems  in the frame of the
quantum Brownian motion as described by Schwinger \cite{Schwinger}
and by Feynman and Vernon \cite{FeynmanVernon}, and from the other
side, the suggestion put forward by 't Hooft in a recent series of
papers \cite{erice,thof1}, according to which Quantum Mechanics
may be an effective theory resulting from a more fundamental
deterministic theory after a process of information loss has taken
place. Some results recently obtained by us point to an intriguing
underlying connection between our approach to dissipative systems,
framed, as said, in the Schwinger and Feynman and Vernon
formalism, and with a strong relation with the thermal field
theory formalism of Takahashi and Umezawa
\cite{UmezawaTaka,Umezawa,dissipation,BGPV,canadian},
and 't Hooft proposal.

In the next section we derive the exact action for a damped
mechanical system (and the special case of the linear oscillator)
from the path integral formulation of the quantum Brownian motion
problem developed by Schwinger and by Feynman and Vernon. We will
closely follow in our discussion refs.\cite{SVW,brownian}.
The doubling of
the phase-space degrees of freedom for dissipative systems and
thermal field theories is discussed and the doubled variables are
related to quantum noise effects. In section 3, the 't Hooft proposal
is discussed and the loss of information due to dissipation in a
classical deterministic system is shown to manifest in the quantum
character of the spectrum of the harmonic oscillator. The
geometric (Berry-Anandan-like) phase arising in dissipative
systems is recognized and related to the zero-point energy of the
quantum oscillator spectrum. The results obtained in
ref.\cite{dissquant} are there reported. In section 4 we consider
the connection with thermal observables recognizing the role
played by the system free energy. Section 5 is devoted to further
remarks and to conclusions.

For the sake of shortness we do not report on the coherent
structure of the states of the quantum system. Details on that
part and on other features of the approach here presented can be
found in the literature (see, e.g.
\cite{dissipation,BGPV,canadian,banerjee}). Also, we do not report about
several applications of our formalism, ranging from the study of
topologically massive gauge theories in the infrared region in
$2+1$ dimensions \cite{BGPV}, the Chern-Simons-like
dynamics of Bloch electrons in solids \cite{BGPV}, the
expanding geometry model in inflationary cosmology \cite{AV1}, to
the study of the quantum brain model
\cite{V,AV2,MyD} and non-commutative geometry \cite{noncomm}.

\section{The exact action for damping}

Our aim in this section is to obtain the exact action for a
particle of mass $m$, damped by a mechanical resistance $\gamma$,
moving in a potential $V$. To be definite, we consider the damped
harmonic oscillator (dho)
\be\lab{1} m \ddot{x}+\gamma\dot{x}+\ka x=0, \ee
as a simple prototype for dissipative systems. Our discussion and
our results also apply, however, to more general systems than the one
represented in (\ref{1}).

The damped oscillator Eq. (\ref{1}) is a non-hamiltonian system
and the canonical formalism, which one needs in order to proceed
to its quantization, cannot be set up \cite{bateman}.
Let us see, however, how one can face the problem by
resorting to well known tools such as the density matrix and the
Wigner function.

We start with the preliminary consideration of the special case of
zero mechanical resistance. The Hamiltonian for an isolated
particle reads
\be\lab{H} H=- \frac{\hbar^2}{2m}\lf(\frac{\partial}{\partial
x}\ri)^2 +V(x). \ee
On the other hand, it is useful to consider the Wigner function,
whose standard expression is, see e.g. \cite{Feynman,Haken},
\be\lab{W} W(p,x,t) = \frac{1}{2\pi \hbar}\int {\psi^* \lf(x -
\frac{1}{2}y,t\ri)\psi \lf(x + \frac{1}{2}y,t\ri)
e^{\lf(-i\frac{py}{\hbar}\ri)}dy} ~, \ee
with the associated density matrix function
\be\lab{8} W(x,y,t)=<x+\frac{1}{2}y|\rho (t)|x-\frac{1}{2}y> =
\psi^* \lf(x-\frac{1}{2}y,t\ri)\psi \lf(x+\frac{1}{2}y,t\ri). \ee

For an isolated particle one obtains the density matrix equation
of motion
\be\lab{5} i\hbar \frac{d \rho}{dt}=[H,\rho ]. \ee
In the coordinate representation, employing
\be\lab{7} x_{\pm}=x\pm \frac{1}{2}y, \ee
Eq. (\ref{5}) reads
\bea \non
&& i\hbar \frac{\partial}{\partial t}<x_+|\rho (t)|x_->=
\\ \lab{6}
&&\lf\{ -\frac{\hbar^2}{2m}\lf[\lf(\frac{\partial}{\partial
x_+}\ri)^2-\lf(\frac{\partial}{\partial x_-}\ri)^2\ri]
+[V(x_+)-V(x_-)] \ri\}<x_+|\rho (t)|x_-> ,
\eea
which, in terms of  $x$ and $y$, is
\be\lab{9a} i\hbar \frac{\partial }{\partial t} W(x,y,t)={\cal
H}_o W(x,y,t) \ee
\be\lab{9b} {\cal H}_o=\frac{1}{m}p_xp_y
+V\lf(x+\frac{1}{2}y\ri)-V\lf(x-\frac{1}{2}y\ri), \ee
\be\lab{9c} p_x=-i\hbar\frac{\partial}{\partial x}, \ \
p_y=-i\hbar\frac{\partial}{\partial y}. \ee
Of course the ``Hamiltonian'' Eq.(\ref{9b}) may be constructed
from the ``Lagrangian''
\be\lab{10}
 {\cal L}_o=m
\dot{x}\dot{y}-V\lf(x+\frac{1}{2}y\ri)+V\lf(x-\frac{1}{2}y\ri).
\ee

Now let us suppose that the particle interacts with a thermal bath
at temperature $T$. The interaction Hamiltonian between the bath
and the particle is taken as
\be\lab{11} H_{int}=-fx, \ee
where $f$ is the random force on the particle at the position $x$
due to the bath.

In the Feynman-Vernon formalism, the effective action for the
particle has the form
\be\lab{12} {\cal A}[x,y]=\int_{t_i}^{t_f}dt\,{\cal
L}_o(\dot{x},\dot{y},x,y) +{\cal I}[x,y], \ee
where ${\cal L}_o$ is defined in Eq.(\ref{10}) and
\be\lab{13} e^{\frac{i}{\hbar}{\cal I}[x,y]}\,=\,
< ( e^{-\frac{i}{\hbar}\int_{t_i}^{t_f}f(t)x_-(t)dt}\, )_-\,
( e^{\frac{i}{\hbar}\int_{t_i}^{t_f}f(t)x_+(t)dt} \,)_+>.
\ee
In Eq.(\ref{13}), the average is with respect to the thermal bath;
``$(.)_{+}$'' and ``$(.)_{-}$'' denote time ordering and anti-time
ordering, respectively; the c-number coordinates $x_{\pm }$ are
defined as in Eq.(\ref{7}). We observe that if the interaction
between the bath and the coordinate $x$ (i.e $H_{int}=-fx$ ) were
turned off, then the operator $f$ of the bath would develop in
time according to $f(t)=e^{iH_\gamma  t/\hbar}fe^{-iH_\gamma
t/\hbar }$ where $H_\gamma$ is the Hamiltonian of the isolated
bath (decoupled from the coordinate $x$). $f(t)$ is the force
operator of the bath to be used in Eq.(\ref{13}).

The reduced density matrix function in Eq.(\ref{8}) for the
particle which first makes contact with the bath at the initial
time $t_i$ is given at a final time by
\be\lab{14} W(x_f,y_f,t_f)=\int_{-\infty}^{\infty
}dx_i\int_{-\infty}^{\infty }dy_i\,
K(x_f,y_f,t_f;x_i,y_i,t_i)W(x_i,y_i,t_i), \ee
with the path integral representation for the evolution kernel
\be\lab{15} K(x_f,y_f,t_f;x_i,y_i,t_i)=
\int_{x(t_i)=x_i}^{x(t_f)=x_f}{\cal D}x(t)
\int_{y(t_i)=y_i}^{y(t_f)=y_f}{\cal D}y(t) \,e^{\frac{i}{\hbar}{\cal
A}[x,y]}. \ee

The evaluation of ${\cal I}[x,y]$ for a linear passive damping
thermal bath requires the use of several Greens functions all of
which have been discussed by Schwinger \cite{Schwinger} and for
shortness here we only mention that the fundamental correlation
function for the random force on the particle due to the thermal
bath is given by (see \cite{SVW})
\be\lab{16} G(t-s)=\frac{i}{\hbar}<f(t)f(s)>. \ee
The retarded and advanced Greens functions are defined by
\be\lab{19a} G_{ret}(t-s)=\theta (t-s)[G(t-s)-G(s-t)], \ee
\be\lab{19b} G_{adv}(t-s)=\theta (s-t)[G(s-t)-G(t-s)]~. \ee
The mechanical impedance $Z(\zeta )$ (analytic in the upper half
complex frequency plane ${\cal I}m ~\zeta >0$) is given by
\be\lab{20} -i\zeta Z(\zeta )=
\int_0^\infty dt \,G_{ret}(t)e^{i\zeta
t}.\ee
and the quantum noise in the fluctuating random force is given by
\be\lab{21} N(t-s)=\frac{1}{2}<f(t)f(s)+f(s)f(t)>, \ee
and is distributed in the frequency domain in accordance with the
Nyquist theorem
\be\lab{22} N(t-s)=\int_0^\infty d\omega \,S_f (\omega)\cos[\omega
(t-s)], \ee
\be\lab{23} S_f(\omega )=\frac{\hbar \omega}{\pi} \coth\frac{\hbar
\omega}{2kT}{\cal R}e Z(\omega +i0^+). \ee
The mechanical resistance is defined by
\be\lab{24} \gamma =lim_{\omega \rightarrow 0}{\cal R}e Z(\omega
+i0^+). \ee

Eq.(\ref{13}) may be now evaluated following Feynman and Vernon
\cite{SVW} as,
\bea \non
{\cal I}[x,y]&=&
\frac{1}{2}\int_{t_i}^{t_f}\int_{t_i}^{t_f}dt ds\,
[G_{ret}(t-s)+G_{adv}(t-s)][x(t)y(s)+x(s)y(t)]
\\ \lab{27}
&&+ \frac{i}{2\hbar}\int_{t_i}^{t_f}\int_{t_i}^{t_f}dt
ds N(t-s)y(t)y(s).
\eea
By defining the retarded force on $y$ and the advanced force on
$x$ as
\bea\lab{28a} F_y^{ret}(t)&=&\int_{t_i}^{t_f}ds
\,G_{ret}(t-s)y(s),
\\ [2mm]
\lab{28b} F_x^{adv}(t)&=&\int_{t_i}^{t_f}ds\,
G_{adv}(t-s)x(s), \eea
respectively, the interaction between the bath and the particle is
then
\bea\nonumber
{\cal
I}[x,y]&=&\frac{1}{2}\int_{t_i}^{t_f}dt\,[x(t)F_y^{ret}(t)+y(t)F_x^{adv}(t)]
\\ \lab{29}
&&+\frac{i}{2\hbar}\int_{t_i}^{t_f}\int_{t_i}^{t_f}dt ds
\,N(t-s)y(t)y(s).
\eea

Thus the real and the imaginary part of the action are finally
given by
\be\lab{30a} {\cal R}e{\cal A}[x,y]=\int_{t_i}^{t_f}dt\,{\cal L},
\ee
\be\lab{30b}
{\cal L}\,=\,m
\dot{x}\dot{y}-\lf[V(x+\frac{1}{2} y)-V(x-\frac{1}{2}y)\ri] +
\frac{1}{2}\lf[x F_y^{ret} + y F_x^{adv}\ri],
\ee
and
\be\lab{30c} {\cal I}m{\cal A}[x,y]=
\frac{1}{2\hbar}\int_{t_i}^{t_f}\int_{t_i}^{t_f}dt ds
\,N(t-s)y(t)y(s).
\ee
respectively. Eqs.(\ref{30a}),(\ref{30b}),(\ref{30c}) are {\it
rigorously exact} for linear passive damping due to the bath when
the path integral Eq.(\ref{15}) is employed for the time
development of the density matrix.

The lesson we learn from our result is that in the classical limit
``$\hbar \rightarrow 0$'' nonzero $y$ yields an ``unlikely
process'' in view of the large imaginary part of the action
implicit in Eq.(\ref{30c}). On the contrary, at quantum level
nonzero $y$ may allow quantum noise effects arising from the
imaginary part of the action \cite{SVW}.

In conclusion, we can consider the approximation to Eq.(\ref{30b})
with $F_y^{ret}=\gamma\dot{y}$ and $F_x^{adv}=-\gamma\dot{x}$,
i.e.
\be\lab{2} {\cal L}(\dot{x},\dot{y},x,y)\,=\,m \dot{x}\dot{y}
-V\lf(x+\frac{1}{2}y\ri)+V\lf(x-\frac{1}{2}y\ri)+\frac{\gamma}{2}(x\dot{y}-
y\dot{x}),
\ee
which implies the classical equations of motion
\bea\lab{3a}
&&m \ddot{x}+\gamma \dot{x}+ \frac{1}{2}\lf[V^\prime
 (x+\frac{1}{2} y)+V^\prime (x-\frac{1}{2} y )\ri]\,=\,0,
\\ \lab{3b}
&& m \ddot{y}-\gamma \dot{y}+ V^\prime
 (x+\frac{1}{2}y )-V^\prime (x-\frac{1}{2}y )\,=\,0.
 \eea

It is easy to see that for $V\lf(x \pm \frac{1}{2}y\ri) =
\frac{1}{2}\ka (x \pm \frac{1}{2}y)^{2}$ Eqs. (\ref{3cx}),(\ref{3cy})
give the dho equation Eq.(\ref{1}) and its complementary equation
for the $y$ coordinate
\bea\lab{3cx} m \ddot{x}+\gamma\dot{x}+\ka x&=&0,
\\ \lab{3cy} m \ddot{y}-\gamma\dot{y}+\ka y&=&0 . \eea
The $y$-oscillator is the time--reversed image of the
$x$-oscillator. If from the manifold of solutions to
Eqs.(\ref{3cx}),(\ref{3cy}) we choose those for which the $y$
coordinate is constrained to be zero, then
Eqs.(\ref{3cx}),(\ref{3cy}) simplify to
\be\lab{3c}  m \ddot{x}+\gamma\dot{x}+\ka x=0,\ \  y=0.  \ee
Thus we obtain a classical damped equation of motion from a
lagrangian theory at the expense of introducing an ``extra''
coordinate $y$, later constrained to vanish. Note that $y(t)=0$ is
a true solution to Eqs.(\ref{3cx}),(\ref{3cy}) (and
(\ref{3a}),(\ref{3b})) so that the constraint is {\it not} in
violation of the equations of motion.

We stress, however, that {\it the role of the ``doubled" $y$
coordinate is absolutely crucial in the quantum regime since there
it accounts for the quantum noise as shown above}. This result
leads us to consider carefully 't Hooft's proposal \cite{erice,thof1}.
When, as customary, one adopts the classical (legitimate) solution
$y = 0$, the $x$ system appears to be open, ``incomplete"; the
loss of information due to dissipation essentially amounts to
neglecting the bath and/or to the ignorance of specific features
of the bath-system interaction, i.e. the ignorance of ``where" and
``how" energy flows out of the system. According to our result,
reverting from the classical level to the quantum level, the loss
of information occurring  at the classical level due to
dissipation manifests itself in terms of ``quantum" noise effects
arising from the imaginary part of the action, to which the $y$
contribution is indeed crucial. In the next sections we analyze in
some details our approach to dissipation in connection with 't
Hooft proposal.

\section{The damped/amplified harmonic oscillator system}

To establish a link with 't~Hooft's quantization scenario it is
important to dwell a bit on some formal aspects  of the
damped--amplified harmonic oscillator  system
(\ref{3cx})--(\ref{3cy}). To do this let us note first that by
defining $x^{\alpha} = (x,y)$ the equations of motion
(\ref{3cx})--(\ref{3cy}) can be written in the compact form
\bea \lab{eqx} m {\ddot{\bf x}} + \ga \sigma_3 {\dot{\bf x}} + \ka
{\bf{x}} = 0 \, .
\eea
This suggest that under appropriate boundary conditions the
$y$--oscillator is the time--reversed image of the
$x$--oscillator. Introducing the metric tensor $g_{\alpha \beta} =
(\sigma_1)_{\alpha \beta}$ the corresponding Lagrangian reads
\bea \label{xylag}
 {\mathcal{L}} =  \frac{m}{2} \
{\dot{\bf{x}}}{\dot{\bf{x}}} + \frac{\ga}{2} \
{\dot{\bf{x}}}\wedge{\bf{x}} -\frac{\kappa}{2} \
{\bf{x}}{\bf{x}}\, ,
%
\eea
with an obvious notation ${\mathbf{a}}{\mathbf{b}} = g_{\alpha
\beta} \ a^{\alpha} b^{\alpha}$ and ${\mathbf{a}}\wedge
{\mathbf{b}} = \varepsilon^{\alpha \beta}\ a_{\alpha} b_{\beta}$
($\varepsilon^{\alpha \beta} = -\varepsilon_{\alpha \beta}$).
It is convenient to reformulate the former
in the rotated coordinate system, i.e.
\begin{eqnarray*}
x_1 = \frac{x+ y}{\sqrt{2}}\, , \,\,\,\,\ x_2 =
\frac{x-y}{\sqrt{2}}\, .
\end{eqnarray*}
In these coordinates the Lagrangian has the form
\begin{eqnarray}
{\mathcal{L}} = \frac{m}{2} \ {\dot{\bf{x}}}{\dot{\bf{x}}} +
\frac{\ga}{2} \ {\bf{x}}\wedge{\dot{\bf{x}}} -\frac{\kappa}{2} \
{\bf{x}}{\bf{x}}\, .
\end{eqnarray}
Here $x^{\alpha} = (x_1,x_2)$ and the metric tensor $g_{\alpha
\beta} = (\sigma_3)_{\alpha \beta}$ (note the change of sign in
the wedge product). Introducing the canonical momenta $p^{\alpha}
=
\partial {\mathcal{L}}/ \partial {\dot{x}_{\alpha}}$  and
$p_{\alpha} \equiv (p_1,p_2)$ we obtain
\begin{equation}
{\mathbf{p}} = m\ \dot{\mathbf{x}} - \frac{\ga}{2} \ \sigma_1
 {\mathbf{x}}\, ,
\end{equation}
and thus the corresponding Hamiltonian reads
\begin{eqnarray}\label{ham1}
H = \frac{1}{2m} \ {\mathbf{p}}^2 + \frac{\gamma}{2m}\
{\mathbf{p}}\wedge {\mathbf{x}} + \frac{1}{2} \left(\kappa -
\frac{\gamma^2}{4m} \right) \ {\mathbf{x}}^2\, .
\end{eqnarray}
The key observation is that with the system (\ref{ham1}) we can
affiliate  the $su(1,1)$ algebraic structure. Indeed, from the
dynamical variables ${{p}}_{\alpha}$ and ${{x}}^{\alpha}$ one may
construct the functions
\begin{eqnarray}
J_1 &=& \frac{1}{2m \Omega} \ p_1 p_2 -  \frac{m \Omega}{2}
\ x_1 x_2\, , \nonumber
\\
J_2 &=& -\frac{1}{2} \ {\mathbf{p}} \wedge {\mathbf{x}} = -
\frac{1}{2} \varepsilon^{\alpha}_{\, \beta} \ p_{\alpha}x^{\beta} =
\frac{1}{2} \left( p_1 x_2 + p_2 x_1 \right)\, , \nonumber
\\
J_3 &=& \frac{1}{4m \Omega} \left(p_1^2 + p_2^2  \right) +
\frac{m\Omega}{4}\ \left(x_1^2  + x_2^2  \right)\, .
\end{eqnarray}
Here $\Om = \sqrt{\frac{1}{m} (\ka-\frac{\ga^2}{4m})}$, with $\ka
>\frac{\ga^2}{4m} $. Applying now the canonical Poisson brackets $\{p_{\alpha},
x^{\beta}\} = g_{\alpha}^{\beta} = \delta_{\alpha}^{\beta}$ we
obtain Poisson's subalgebra
\begin{eqnarray}
\{ J_2,J_3\} = J_1\, , \,\,\,\, \{ J_3, J_1 \} = J_2\, , \,\,\,\,
\{ J_1,J_2 \} = - J_3\, . \label{poissons}
\end{eqnarray}
The algebraic structure (\ref{poissons}) corresponds to $su(1,1)$
algebra \cite{wib}. The quadratic
Casimir for the algebra (\ref{poissons}) is defined as \cite{wib}
\begin{eqnarray}
C = \frac{1}{2} \ \left( J_1^2 + J_2^2 - J_3^2 \right) =
-\frac{1}{2} \left( \frac{{\mathbf{p}}^2 + m^2
\Omega^2 }{4m\Omega}  \right)^2 \equiv - \frac{1}{2} \
{\mathcal{C}}^2\, .
\end{eqnarray}
In terms of $J_2$ and the Casimir ${\mathcal{C}}$ the
Hamiltonian (\ref{ham1}) can be formulated as
\begin{eqnarray}\lab{ham2}
H \ = \ 2  \left(\Omega {\mathcal{C}} - \Gamma J_2   \right)\, ,
\end{eqnarray}
with $\Ga = \ga/ 2 m$. It might be shown \cite{BGPV} that when
(\ref{ham1}) is quantized then $SU(1,1)$ is the dynamical group of
the system. A formal simplification occurs when the hyperbolic
coordinates are introduced, i.e.
\bea
 &&x_1 \ =\  r \cosh u \, , \,\,\,\,
x_2 \ = \ r \sinh u \,  . \eea
Then $J_2$ and ${\mathcal{C}}$ have a particularly simple
structure \cite{BGPV}, namely
\bea\label{pu}
{\mathcal{C}}  \ = \ \frac{1}{4 \Om m}\lf[ p_r^2 -
\frac{1}{r^2}p_u^2 + m^2\Om^2 r^2\ri]\, , \,\,\,\,  J_2 \ = \
\frac{1}{2}\, p_u\, , \eea
Let us finally note that the dynamical system described by the
Lagrangian (\ref{xylag}) is sometimes
called Bateman's dual system \cite{bateman}.

\section{Deterministic dissipative systems and quantization}
In a recent series of papers  \cite{erice,thof1}, G.'t Hooft has
put forward the idea that Quantum Mechanics may result from a more
fundamental deterministic theory, after that a process of information
loss has taken place. He has found a class of Hamiltonian systems
which remain deterministic, even when they are described by means
of Hilbert space techniques. The truly quantum systems are
obtained when constraints are imposed on the original Hilbert
space: these constraints implement the information loss.

More specifically, the Hamiltonian for such systems is of the form
\bea\label{thofH} H = \sum_{i}p_{i}\, f_{i}(q)\,,
\eea
where $f_{i}(q)$ are non--singular functions of the canonical
coordinates $q_{i}$. The crucial point is that equations for the
$q$'s (i.e. $\dot{q_{i}} = \{q_{i}, H\} = f_{i}(q)$) are decoupled
from the conjugate momenta $p_i$ and this implies \cite{erice}
that the system can be described deterministically even when
expressed in terms of operators acting on the Hilbert space. The
condition for the deterministic description is the existence of a
complete set of observables commuting at all times, called {\em
beables} \cite{Bell:1987hh} - a condition which is guaranteed for
the systems of Eq.(\ref{thofH}), for which such a set is given by
the $q_i(t)$ \cite{erice}.

A problem with the above mentioned class of Hamiltonians is that
they are not bounded from below. This might be cured by splitting
$H$ in Eq.(\ref{thofH}) as \cite{erice}:
\bea && H = H_1 - H_2\quad,\quad H_1 = \frac{1}{4\rho}\left( \rho
+ H\right)^{2}\;\; ,\;\;\; H_2 = \frac{1}{4\rho}\left( \rho -
H\right)^{2}\, ,  \label{1.5} \eea
with  $\rho$ a certain time--independent, positive function of
$q_{i}$. As a result, $H_1$ and $H_2$ are positively
(semi)definite and $ \{H_1, H_2\} = \{\rho , H \} =  0\,$ .

To get the lower bound for the Hamiltonian one thus imposes the
constraint condition onto the Hilbert space:
\bea H_2|\psi \ran = 0\, , \label{3.8} \eea
which projects out the states responsible for the negative part of
the spectrum. In the deterministic language this means that one
gets rid of the unstable trajectories \cite{erice}.
In the line of 't Hooft's
proposal, it has been shown \cite{Elze} that a
reparametrization-invariant time technique in a specific model
also leads to a quantum dynamics emerging from a deterministic
classical evolution. Deterministic models with discrete time evolution
have been recently studied  in Ref.\cite{contraction}.

We now show that the above discussed system of damped-antidamped
oscillators does provide indeed an explicit realization of 't
Hooft mechanism. Furthermore, we shall see that there is a
connection between the zero point energy of the
quantum harmonic oscillator and the geometric phase
of the (deterministic) system of damped/antidamped oscillators.

The first thing we need to realize is   that the
Hamiltonian of our model belongs to the same class of the
Hamiltonians considered by 't Hooft. Indeed Eq.(\ref{ham2}) can
be then rewritten as \cite{dissquant}:
\bea \lab{pqham} H &=& \sum_{i=1}^2p_{i}\, f_{i}(q)\,, \eea
with $f_1(q)=2\Om$,  $f_2(q)=-2\Ga$, provided we use the canonical
transformation:
\begin{eqnarray}
&&q_{1} = \int \frac{dz \;\; m\Omega}{\sqrt{4 J_{2}^{2} + 4m\Omega
{\cal{C}
} z - m^{2}\Omega^{2} z^{2}}}\, ,\\
&&q_{2} = 2u + \int \frac{dz}{z} \, \frac{2J_{2}}{\sqrt{4
J_{2}^{2} + 4m\Omega {\cal{C}}
z - m^{2}\Omega^{2} z^{2}}}\, ,\\
&&p_1 = {\cal C}\, , \;\;\;\;\ p_2 = J_2 \, , \label{can1}
\end{eqnarray}
with $z=r^2$. One has $\{q_{i},p_i\} =1$, and the other Poisson
brackets vanishing. Thus  $J_2$ and ${\cal C}$  are beables. Yet
also $q_1$ and $q_2$ are beables
 as it can be directly seen from the
Hamiltonian (\ref{pqham}).

Now we set
\bea
&&H = H_\IJ - H_{\II} \quad, \quad
H_\IJ = \frac{1}{2 \Om {\cal C}} (2 \Om {\cal C} - \Ga
J_2)^2\;\;\; ,\;\;\;\; H_{\II} = \frac{\Ga^2}{2 \Om {\cal C}}
J_2^2\, .  \label{split} \eea
Of course, only  nonzero $r^{2}$ should be taken into account in
order for $\cal C$ to be invertible. Note that ${\cal C}$ is a
constant of motion (being the Casimir operator): this ensures that
once it has been chosen to be positive, as we do from now on, it
will remain such at all times.

We  then implement the constraint
\bea \label{thermalcondition}\quad J_2 |\psi\ran = 0\, , \eea
which defines the physical states. Although the system
(\ref{pqham}), i.e.(\ref{ham2}), is deterministic, $|\psi\ran$ is
not an eigenvector of $u$ ($u$ does not commute with $p_u)$. Of
course, if one does not use the operatorial formalism to describe
our system, then $p_u=0$ implies $u = -\frac{\ga}{2 m} t$.
 Eq.(\ref{thermalcondition})
implies
\bea \lab{17} H |\psi\ran= H_\IJ |\psi\ran=  2\Om {\cal
C}|\psi\ran = \left( \frac{1}{2m}p_{r}^{2} +
\frac{K}{2}r^{2}\right) |\psi \ran \, , \eea
where  $K\equiv m \Om^2$. $H_\IJ$ thus reduces to the Hamiltonian
for the linear harmonic oscillator $\ddot{r} + \Om^2 r =0 $. The
physical states are even with respect to time-reversal
($|\psi(t)\ran  = |\psi(-t)\ran$) and periodical with period $\tau
= \frac{2\pi}{\Omega}$.

Having denoted with $|\psi\ran$ the physical states, we now introduce the
states $|\psi(t)\ran_{H}$  and $|\psi(t)\ran_{H_\IJ}$ satisfying
the equations:
\bea \lab{S1} i \hbar \frac{d}{dt} |\psi(t)\ran_{H} &=& H
\,|\psi(t)\ran_{H}~,
\\ \lab{S2}
i \hbar \frac{d}{dt} |\psi(t)\ran_{H_\IJ} &= &2 \Om {\cal C}
|\psi(t)\ran_{H_\IJ} \, . \eea
Eq.(\ref{S2}) describes the 2D ``isotropic'' (or ``radial'')
harmonic oscillator. $ H_\IJ = 2 \Om{\cal C} $ has the  spectrum
${\cal H}^n_\IJ= \hbar \Om n$, $n = 0, \pm 1, \pm 2, ...$.
According to our choice for ${\cal C}$ to be positive, only
positive values of $n$ will be considered.

The generic state $|\psi(t)\ran_{H}$ can be written as
\bea\lab{eqt0} |\psi(t)\ran_{H} = {\hat{T}}\lf[ \exp\lf(
\frac{i}{\hbar}\int_{t_0}^t 2 \Ga J_2 dt' \ri) \ri]
|\psi(t)\ran_{H_\IJ} ~, \eea
where ${\hat{T}}$ denotes time-ordering. Note that here
$\hbar$ is introduced on purely dimensional grounds and its
actual value cannot be fixed by the present analysis.

\begin{figure}[b]
\begin{center}
\includegraphics[width=.8\textwidth]{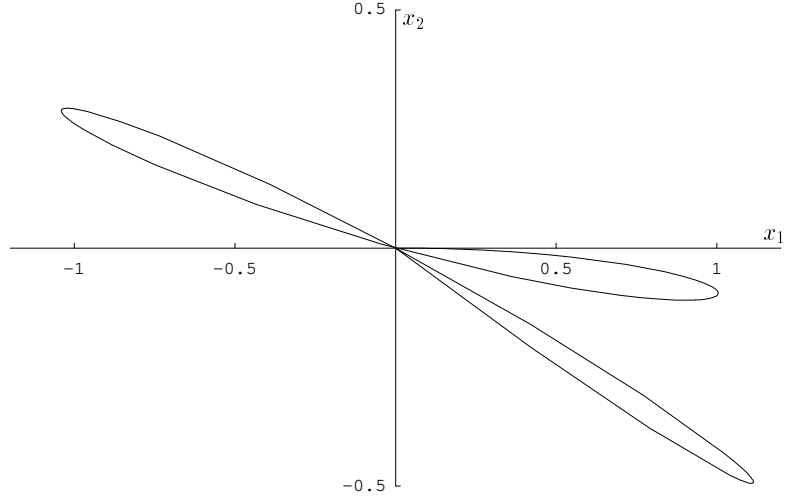}
\end{center}
\caption[]{Trajectories for $r_0=0$ and $v_0=\Om$, after three
half-periods for $\ka =20$, $\ga=1.2$ and $m=5$. The ratio
$\int_0^{\tau/2}({\dot x}_1 x_2 - {\dot x}_2 x_1)dt /{\cal E}
=\pi\frac{\Ga}{m\Om^3}$  is preserved. ${\cal E}$ is the initial  energy:
${\cal
E}=\frac{1}{2} m v_0^2 + \frac{1}{2} m \Om^2 r_0^2$.}
\label{eps1}
\end{figure}

We obtain \cite{dissquant}:
\bea &&\,_{H}\lan \psi(\tau) | \psi(0) \ran_{H} =\,_{H_\IJ}\!\lan
\psi(0)| \exp\lf( i \int_{C_{0\tau}} A(t') dt' \ri)
|\psi(0)\ran_{H_\IJ} \equiv e^{i \phi} \, , \label{berryphase}
\eea
where the contour $C_{0\tau}$ is the one going from $t'=0$ to
$t'=\tau$ and back and  $A(t) \equiv \frac{\Ga m}{ \hbar}({\dot x}_1 x_2 - {\dot
x}_2 x_1)$. Note that $({\dot x}_1 x_2 - {\dot x}_2 x_1) dt$ is the area
element in the $(x_1,x_2)$ plane enclosed by the trajectories (see
Fig.1).
Notice also that the
evolution (or dynamical) part of the phase does not enter in
$\phi$, as the integral in Eq.(\ref{berryphase}) picks up  a
purely geometric contribution \cite{Berry}.

Because the physical states $|\psi\rangle$ are periodic ones, let
us focus our attention on those. Following \cite{Berry}, one may
generally write
\begin{eqnarray}
|\psi(\tau)\ran &= &e^{ i\phi -
\frac{i}{\hbar}\int_{0}^{\tau}\langle \psi(t)| H |\psi(t) \rangle
dt} |\psi(0)\ran \,=\, e^{- i2\pi n} | \psi(0)\ran \, ,
\label{eqtH}
\end{eqnarray}
i.e. $\frac{ \langle \psi(\tau)| H |\psi(\tau) \rangle  }{\hbar}
\tau - \phi = 2\pi n$, $n = 0,  1,  2, \ldots $,  which by using
$\tau = \frac{2 \pi}{\Om}$ and $\phi = \alpha \pi$, gives
\bea\lab{spectrum} {\cal H}_{{_I},e\!f\!f}^n \equiv \langle
\psi_{n}(\tau)| H |\psi_{n}(\tau) \rangle= \hbar \Om \lf( n +
\frac{\alpha}{2} \ri) ~, \eea
where the index $n$ has been introduced to exhibit the $n$
dependence of the state  and  the corresponding energy. ${\cal
H}_{\I,e\!f\!f}^n$ gives the effective $n$th energy level of the
physical system, namely the energy given by ${\cal H}_{\I}^n$
corrected by its interaction with the environment. We thus see
that the dissipation term $J_2$ of the Hamiltonian  is actually
responsible for the ``zero point energy" ($n = 0$): $E_{0}
=\frac{\hbar}{2} \Om \alpha$.

As well known, the zero point energy is the ``signature" of
quantization since in Quantum Mechanics it is formally due to the
non-zero commutator of the canonically conjugate $q$ and $p$
operators. Thus dissipation manifests itself as ``quantization".
In other words, $E_0$, which appears as the ``quantum
contribution" to the spectrum of the conservative evolution of
physical states, signals the underlying dissipative dynamics.  If
we want to match the Quantum Mechanics zero point energy, we have
to fix $\alpha = 1$, which gives \cite{dissquant} $\Om =
\frac{\ga}{m}$.

\section{Thermodynamics}

In order to better understand the dynamical role of $J_2$ we
rewrite Eq.(\ref{eqt0}) as
\bea\lab{eqt0U} |\psi(t)\ran_{H} = {\hat T}\lf[ \exp\lf(i
\frac{1}{\hbar} \int_{u(t_0)}^{u(t)} 2 J_2 du'\ri) \ri]
|\psi(t)\ran_{H_\IJ} \, , \eea
by using $u(t) = - \Ga t$. Accordingly, we have
\bea \lab{Su} -i \hbar \frac{\pa}{\pa u} |\psi(t)\ran_{H} = 2J_{2}
|\psi(t)\ran_{H} \, . \eea
We thus see that $2 J_2$ is responsible for shifts (translations)
in the $u$ variable, as is to be expected since $2 J_{2} = p_{u}$
(cf. Eq.(\ref{pu})). In operatorial notation we can write indeed
$p_{u} =- i \hbar \frac{\pa}{\pa u}$. Then, in full generality,
Eq.(\ref{thermalcondition}) defines families of physical states,
representing stable, periodic trajectories (cf. Eq.(\ref{17})). $2
J_{2}$ implements transition from family to family, according to
Eq.(\ref{Su}). Eq.(\ref{S1}) can be then rewritten as
\bea \lab{S11} i \hbar \frac{d}{dt} |\psi(t)\ran_{H} = i \hbar
\frac{\pa}{\pa t} |\psi(t)\ran_{H} + i \hbar
\frac{du}{dt}\frac{\pa}{\pa u} |\psi(t)\ran_{H}\, , \eea
where the first term on the r.h.s. denotes of course derivative
with respect to the explicit time dependence of the state. The
dissipation contribution to the energy is thus described by the
``translations" in the $u$ variable.  It is then interesting to
consider the derivative
\bea\lab{T} \frac{\pa S}{\pa U} = \frac{1}{T}\,. \eea
From Eq.(\ref{pqham}), by using $S \equiv \frac{2 J_{2}}{\hbar}$
and $U \equiv  2 \Om {\cal C}$, we obtain $T = \hbar \Ga$. Eq.
(\ref{T}) is the defining relation for temperature in
thermodynamics
 (with $k_B = 1$) so that one could formally regard $\hbar
\Ga$ (which dimensionally is an energy) as the temperature,
provided the dimensionless quantity $S$ is identified with the
entropy. In such a case, the ``full Hamiltonian'' Eq.(\ref{pqham})
plays the role of the free energy ${\cal F}$: $H = 2 \Om {\cal
C} - (\hbar \Ga) \frac{2 J_2}{\hbar} = U - TS = {\cal F}$.  Thus
$2 \Ga J_{2}$ represents the heat contribution in $H$ (or $\cal
F$). Of course, consistently, $\lf. \frac{\pa {\cal F} }{\pa T
}\ri|_\Om = - \frac{2 J_2}{\hbar}$.  In conclusion $\frac{2
J_{2}}{\hbar}$ behaves as the entropy, which is not surprising
since it controls the dissipative (thus irreversible) part of the
dynamics.

We can also take the derivative of ${\cal F}$ (keeping $T$ fixed)
with respect to $\Om$. We then have
\bea \lf. \frac{\pa {\cal F}}{\pa \Om}\ri|_T  = \lf. \frac{\pa
U}{\pa \Om}\ri|_T  =m r^2 \Om\, , \eea
which is the angular momentum: this is to be expected since it is
the conjugate variable of the angular velocity $\Om$. It is also
suggestive that the temperature $\hbar \Ga$ is actually given by
the background zero point energy: $\hbar \Ga = \frac{\hbar
\Om}{2}$.

In the light of the above results, the condition
(\ref{thermalcondition}) can be then interpreted as a condition
for  an adiabatic physical system. $\frac{2 J_{2}}{\hbar}$ might
be viewed as an analogue of the Kolmogorov--Sinai entropy for
chaotic dynamical systems.

\section{Conclusions}

In this report we have  reviewed
some aspects of the quantization of the damped harmonic
oscillator.

In the framework of the path integral formulation
developed by Schwinger and by Feynman and Vernon, we have  discussed
the doubling of the phase-space degrees of freedom for dissipative
systems and thermal field theories. We have shown how  the doubled
variables are related to quantum noise effects.

We have then discussed some algebraic features of the system of
damped-antidamped harmonic oscillators which allows for a canonical
treatment of quantum dissipation.

We also considered the relation of this model with the
't Hooft proposal,
according to which the loss of information due to dissipation in a
classical deterministic system manifests itself in the quantum
features of the system. We have shown that the quantum spectrum of
the harmonic oscillator can be obtained from the
dissipative character of the underlying deterministic
system.

Finally, we have discussed the thermodynamical features of our system.

\bigskip
\bigskip

{\bf Acknowledgments}
\smallskip

We would like to thank the organizers of the Piombino workshop
DICE2002 on ``Decoherence, Information, Complexity and Entropy''
where some of the results contained in this report have been
presented. We also thank the ESF network COSLAB, MIUR, INFN, INFM
and  EPSRC for partial support.

%


\begin{thebibliography}{8.}
\addcontentsline{toc}{section}{References}


\bibitem{Schwinger} J. Schwinger:  J. Math. Phys. \textbf{2}
(1961), 407.

\bibitem{FeynmanVernon}
R.P. Feynman, F.L. Vernon:  Annals Phys. \textbf{24}, 118 (1963).

\bibitem{erice} G.~'t Hooft: in
{\em ``Basics and Highlights of Fundamental Physics''}, Erice,
(1999) [hep-th/0003005].

\bibitem{thof1}
G.~'t Hooft: [hep-th/0104080]; [hep-th/0105105]; [quant-ph/0212095].


\bibitem{UmezawaTaka}
Y. Takahashi, H. Umezawa: Collective Phenomena
\textbf{2}, 55 (1975).

\bibitem{Umezawa} H.Umezawa: \emph{Advanced field theory: micro,
macro and thermal concepts} (American Institute of Physics, N.Y.
1993);
\\
H. Umezawa, M. Matsumoto, M. Tachiki:
\emph{Thermo Field Dynamics
and Condensed States} (North-Holland, Amsterdam, 1982).

\bibitem{dissipation}
E.~Celeghini, M.~Rasetti, G.~Vitiello:
Annals Phys.  \textbf{215}, 156 (1992).


\bibitem{BGPV}
M.~Blasone, E.~Graziano, O.~K.~Pashaev, G.~Vitiello:
Annals Phys.  \textbf{252} 115, (1996).

\bibitem{canadian}
M.~Blasone, P.~Jizba:
Can. J. Phys. \textbf{80}, 645 (2002); [quant-ph/0102128].

\bibitem{SVW}
Y.~N.~Srivastava, G.~Vitiello, A.~Widom:
Annals Phys. \textbf{238}, 200 (1995).

\bibitem{brownian}
M.~Blasone, Y.~N.~Srivastava, G.~Vitiello, A.~Widom:
Annals Phys. \textbf{267}, 61 (1998).



\bibitem{dissquant}
M.~Blasone, P.~Jizba, G.~Vitiello:
Phys. Lett. A \textbf{287}, 205 (2001).


\bibitem{banerjee}
R.~Banerjee:
Mod. Phys Lett. A \textbf{17}, 631 (2002);
R.~Banerjee, P.~Mukherjee: J. Phys. A \textbf{35}, 5591 (2002).



\bibitem{AV1}
E.~Alfinito, G.~Vitiello:
 Class. Quant. Grav. \textbf{17}, 93 (2000).

\bibitem{V} G.~Vitiello:
 Int. J. Mod. Phys. \textbf{B9}, 973 (1995).


\bibitem{AV2} E.~ Alfinito,  G.~Vitiello:
 Int. J. Mod. Phys. \textbf{B14}, 853 (2000).

\bibitem{MyD} G.~Vitiello:
\emph{My Double Unveiled}, (John Benjamins, Amsterdam 2001).


\bibitem{noncomm}
S.~Sivasubramanian, Y.~N.~Srivastava, G.~Vitiello, A.~Widom:
[quant-ph/0301005].

\bibitem{bateman}
H.~Bateman:  Phys. Rev. \textbf{38}, 815 (1931);
\\
P.~M.~Morse, H.~Feshbach:
\emph{Methods of Theoretical Physics}, Vol.I, pag. 298
(McGraw--Hill, New York, 1953);
\\
H.~Dekker: Phys.\ Rept.\  \textbf{80}, 1 (1981).


\bibitem{Feynman} R.~P.~Feynman: \emph{Statistical Mechanics},
(The Benjamin/Cummings Publ. Co., INC., Reading, Massachusetts,
1972).

\bibitem{Haken} H.~Haken: \emph{Laser Theory} Springer-Verlag, Berlin 1984.



\bibitem{wib}
B.~G.~Wybourne: \emph{Classical Groups in Physics}, (John Wiley \&
Sons, Inc., London, 1974).


\bibitem{Bell:1987hh}
J.~S.~Bell, \emph{Speakable and unspeakable in Quantum Mechanics}
(Cambridge University Press, 1987).

\bibitem{Elze}
H.~T.~Elze, O.~Schipper:
Phys. Rev. D \textbf{66}, 044020 (2002).

\bibitem{contraction}
M.~Blasone, E.~Celeghini, P.~Jizba, G.~Vitiello:
[quant-ph/0208012].



\bibitem{Berry}
J.~Anandan, Y.~Aharonov,
Phys. Rev. Lett.  \textbf{65}, 1697 (1990).

\end{thebibliography}
\end{document}